\documentclass[traditabstract]{aa}
\usepackage{graphicx}
\usepackage{natbib}
\bibpunct{(}{)}{;}{a}{}{,}

\newcommand{\saxj}{SAX J1808.4--3658}
\newcommand{\xmm}{\it XMM-Newton}
\def\ltsima{$\; \buildrel < \over \sim \;$}
\def\simlt{\lower.5ex\hbox{\ltsima}}
\def\gtsima{$\; \buildrel > \over \sim \;$}
\def\simgt{\lower.5ex\hbox{\gtsima}}

\begin{document}

\title{Timing of the 2008 outburst of {\saxj} with XMM-Newton:
A stable orbital-period derivative over ten years}

\author{L.~Burderi$^{1}$\thanks{mail to burderi@dsf.unica.it} \and 
A.~Riggio$^{1}$ \and T.~Di Salvo$^{2}$ \and A.~Papitto$^{3,4}$ \and 
M.~T.~Menna$^{4}$ \and A.~D'A\`i$^{2}$ \and R.~Iaria$^{2}$
}

\institute{Dipartimento di Fisica, Universit\`a degli Studi di Cagliari, 
SP Monserrato-Sestu, KM 0.7, Monserrato, 09042 Italy
\and Dipartimento di Scienze Fisiche ed Astronomiche, Universit\`a di Palermo, 
via Archirafi 36, Palermo, 90123, Italy 
\and Dipartimento di Fisica, Universit\`a degli Studi di Roma 'Tor Vergata', 
via della Ricerca Scientifica 1,00133 Roma,Italy 
\and INAF Osservatorio Astronomico di Roma, via Frascati 33, Monteporzio Catone, 
00040, Italy  }

\abstract
{We report on a timing analysis performed on a 62-ks long {\xmm} 
observation of the accreting millisecond pulsar {\saxj} during the 
latest X-ray outburst that started on September 21, 2008.
By connecting the time of arrivals of the pulses observed during 
the {\xmm} observation, we derived the best-fit orbital solution 
and a best-fit value of the spin period for the 2008 outburst. 
Comparing this new set of orbital parameters and, in particular, 
the value of the time of ascending-node passage with the orbital
parameters derived for the previous four X-ray outbursts of {\saxj}
observed by the PCA onboard {\it RXTE}, we find an updated 
value of the orbital period derivative, which turns out to be 
$\dot P_{\rm orb} = (3.89 \pm 0.15) \times 10^{-12}$ s/s.
This new value of the orbital period derivative agrees
with the  previously reported value, 
demonstrating that the orbital period 
derivative in this source has remained stable over the past ten years. 
Although this timespan is not sufficient yet for confirming the secular
evolution of the system, 
we again propose an explanation of this behavior in terms
of a highly non-conservative mass transfer 
in this system, where the accreted mass (as derived from the X-ray 
luminosity during outbursts) accounts for a mere 1\% of the mass 
lost by the companion.}

\titlerunning{A Stable Orbital Period Derivative over the Last Ten Years}
\authorrunning{L. Burderi et al.}

\maketitle

\section{Introduction}

Accreting millisecond pulsars (AMSP) are generally interpreted as the
evolutionary link between low-mass X-ray binaries and rotation-powered 
millisecond radio pulsars. 
According to the recycling scenario, the last are
in fact formed after a phase of mass accretion from a low-mass companion star,
which ultimately spins the neutron star up to ms periods
\citep{Alpar_82}. 
{\saxj} was the first AMSP discovered \citep{Wijnands_98}. 
In ten years the population of AMSP has grown to ten sources, 
all residing in close and
transient binaries, but {\saxj} can still be considered the cornerstone of its
class as it has repeatedly gone in outburst almost every two years after its
first detection, making it the most observationally rich source of its class.
Its timing and orbital properties have been extensively studied; while the 
timing behavior of the spin frequency is still debated because of the presence
of timing noise in the phase delays (see \citealt{Burderi_06}, B06 hereinafter; 
\citealt{Hartman_08}, H08), there is  good agreement as regards the 
orbital parameters of this source, which are now known with extreme precision 
(H08; \citealt{DiSalvo_08}, DS08). 

In particular, a timing of the past four outbursts of {\saxj} observed with the 
PCA onboard the {\it RXTE} satellite (spanning more than 7 years from April 
1998 to October 2005) has allowed  the orbital period of this source  to be derived
with  relative uncertainty 
$\Delta P_{orb} / P_{orb}$ of $\sim 2 \times 10^{-9}$ (dS08, H08).
More interesting, the orbital period of the source shows a significant 
derivative, which indicates that it is increasing at a large, as well as 
puzzling, rate of $\dot P_{\rm orb} = (3.5 \pm 0.2) \times 10^{-12}$ s/s.
The value of the orbital period derivative is at least
one order of magnitude greater than what is expected for conservative mass
transfer driven by gravitational radiation (GR; DS08, H08).

To continue the monitoring of the orbital period derivative
and to further improve the orbital solution for this interesting system,
we analyzed a  62 ks  {\xmm} Target  of  Opportunity (ToO)
observation  of this  source, performed during  the latest  outburst  
in October 2008.

\section{Timing analysis and results}

{\saxj} was found in outburst on 2008 September 21 by {\it RXTE} , and
since then it has been the object of an intensive observational campaign.
{\xmm} observed {\saxj} as a ToO observation for $62$ ks on 2008 October 1
(start time MJD 54739.99517), roughly one week after the assumed
outburst peak. In this paper we concentrate on the pn data, which have the 
best statistics for timing analysis. A detailed description of the {\xmm} 
observation and data reduction has already been reported in \citet{Papitto_08}.
The arrival times of all the pn events were referred to the solar system 
barycenter by using the {\tt barycen} tool in SAS~v.8.0.0. As the best 
estimate for the source coordinates, we considered the position of the radio counterpart, 
which has an uncertainty of 0.4 arcsec (90\% c.l., \citealt{Rupen_02}) 
and is compatible with the optical counterpart \citep{Giles_99}.

We used the {\xmm}/pn data to improve the orbital solution for {\saxj}.
To this aim we also re-analyzed the previous four outbursts of
this source observed by RXTE/PCA, namely 
1998 April (Obs.\ ID P30411), 2000 February (Obs.\ ID P40035), 
2002 October (Obs.\ ID P70080), and 2005 June (Obs.\ ID P91056 and 
Obs.\ ID P91418) outbursts, respectively (see 
H08 for a detailed description of these observations).
With this new analysis we present a more conservative treatment of the errors
on the phase delays, to avoid any possible underestimation of the 
uncertainties on the orbital parameters caused by the phase changes and shifts
that are known to occur during outbursts.

\begin{figure}
\hspace{-1cm}
\includegraphics[angle=-90,width=10.0cm]{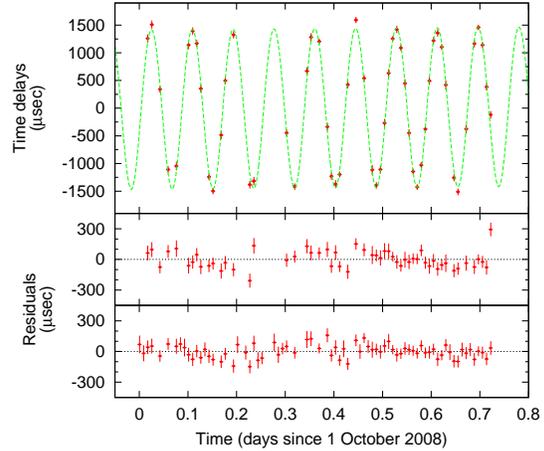}
\vspace{-1cm}
\caption{
Phase delays (in $\mu$s) {\it vs.} time during the {\xmm} observation after the 
first iteration. To show for clarity the orbital correction against an 
horizontal line, the frequency adopted for the folding is the best-fit spin 
frequency of the 2008 outburst. 
The orbital modulation is clearly visible (top panel) and is well-fitted 
by our orbital solution, as can be seen in the post-fit residuals (middle and
bottom panels).}
\label{fig:fig2}
\end{figure}

For each  outburst, we repeat the analysis performed by DS08,
with a few differences.
Differential corrections to the adopted orbital parameters were 
found by fitting the pulse phases as a function of time for each 
outburst (see {\it e.g.} Fig~\ref{fig:fig2}). 
In general, any residual orbital modulation is superposed onto a 
long-term variation in the phases, e.g.\ caused by a variation of the
spin. However, 
\saxj\ shows very complex behavior of the pulse phases with time, with 
phase shifts, probably caused by variations in the pulse shape, that are
difficult to interpret (see e.g.\ B06, H08). To model this complex behavior of 
the pulse phases with time, we fit the phases of 
each of these intervals with the formula for the differential corrections 
to the orbital parameters (see e.g.\ \citealt{Deeter_81}) 
plus a polynomial (up to $8^{\rm th}$ degree, depending
on the irregular behavior of the pulse phases). This technique
is complementary to the technique adopted in dS08, where     
we preferred to restrict the fit of the differential corrections to the 
orbital parameters to intervals in which the long-term variation and/or 
phase shifts were negligible and considered consecutive intervals with 
a duration of at least 4 orbital periods (depending on the statistics), 
fitting the phases of each of these intervals only with the formula for the 
differential corrections of the orbital parameters. We have verified that the 
two techniques give similar orbital corrections, but we now obtain somewhat
larger uncertainties on the derived orbital parameters. 

Over the 10-year base line {\saxj} has been observed, the
measured $\dot{P}_{\mathrm{orb}}$ should have produced  $\approx 1$
ms change in $P_{\mathrm{orb}}$.  However, the uncertainty on the
individual $P_{\mathrm{orb}}$ determinations from the local orbital
solution during each outburst are about 0.5--2.0 ms (see also H08);
accordingly, no significant corrections to $P_{\mathrm{orb}}$ have
been directly observed.  Thus we fix our parameter
$P_{\mathrm{orb,0}} = 7249.156499$ s (dS08) and  improve its
measure below using the same technique as in dS08.
For each of the five outbursts, we find similar corrections for 
the projected semimajor axis of the neutron star orbit 
and upper limits on the eccentricity.
We hence combine the measurements corresponding to each of the outbursts 
by computing the error-weighted means of the corrections to the projected 
semimajor axis 
and of the upper limits on the eccentricity.
On the other hand, we find that the correction 
to the predicted times of
passage of the NS at the ascending node at the beginning of each outburst,
$T^*_{m \; {\rm predicted}} = T^*_0 + N P_{\rm orb\; 0}$, 
is different among the five outbursts.
Here $T^*_0$ is the time of ascending node passage given by dS08 
\footnote{The actual $T^*_0$ used in this work is
the $T^*_0$ of dS08 decremented by $P_{\rm orb \; 0}$ in order to 
have the time of ascending node passage just before the beginning of RXTE data
of the 1998 outburst which occurred at $T_0 = 50914.8099$ MJD.},
and the integer $N$ is the exact number of orbital cycles elapsed between two 
different ascending node passages; i.e., \ $N$ is the closest integer to 
$(T^*_m - T^*_0)/ P_{\rm orb \;0}$ under the assumption that 
$| T^*_m -  T^*_{m \; {\rm predicted}} |<< P_{\rm orb \; 0}$ 
that we have verified {\it a posteriori}, and $m = 1998, 2000, 2002, 2005$, 
and $2008$.
These points show a clear parabolic trend that we fitted to the formula
\begin{equation}
\delta T^*_m = \delta T^* 
+ \delta P_{\rm orb} \times N + (1/2) 
\dot P_{\rm orb} P_{\rm orb \; 0} \times N^2
\label{eq1}
\end{equation}
where $\delta T^*$, $\delta P_{\rm orb}$, 
and $\dot P_{\rm orb}$ (the orbital period derivative)
are the fit parameters.
To compute the correct errors on the derived parameters, 
we also considered the errors induced by the uncertainties in the time coordinate 
(which we assumed to be half of the duration of each outburst) following the 
standard procedure for the propagation of these errors (see e.g.\ 
\citealt{Bevington_03}).

We iterate the method described above until the upper limit on the eccentricity, 
as well as the corrections on the projected semimajor axis, the orbital period, 
and its derivative 
were all compatible with zero within the computed errors. 
In Fig.~\ref{fig:fig2} we plot the phase delays obtained 
for the {\xmm}/pn data with the orbital solution
of DS08 neglecting the orbital-period derivative. The sinusoidal oscillation
is clearly visible (top panel) and mainly stems from  the $\sim 35$ s delay in 
the ascending node passage with respect to the one predicted with a constant 
orbital period. In the middle panel, we show the time residuals after our first 
improvement of the orbital solution (reduced $\chi^2 = 1.86$ for 49 d.o.f.).
In the bottom panel, the time residuals are shown with respect to the final orbital 
solution shown in Table~\ref{tabps} (reduced $\chi^2 = 1.07$ for 72 d.o.f.). 
In this case more points are visible in the residuals. This is
caused by how improving the orbital solution results in a higher
statistical significance for each point, making acceptable phase-points that 
were rejected with our previous solution.
In Fig.~\ref{fig:fig4}
we plot the final $\delta T^*_m$'s vs time, together with the best-fit 
parabola. The fit was good with a reduced $\chi^2 = 1.55$ (for 2 d.o.f.). 
This corresponds to a probability of $21\%$ of obtaining a $\chi^2$ that is larger 
than the one we found. 
Our result is therefore acceptable, since the probability 
we obtained is well above the conventionally accepted significance level of  
$5\%$ (see e.g.\ {\citealt{Bevington_03}).
\begin{figure}
\hspace{-1cm}
\includegraphics[angle=-90,width=10.0cm]{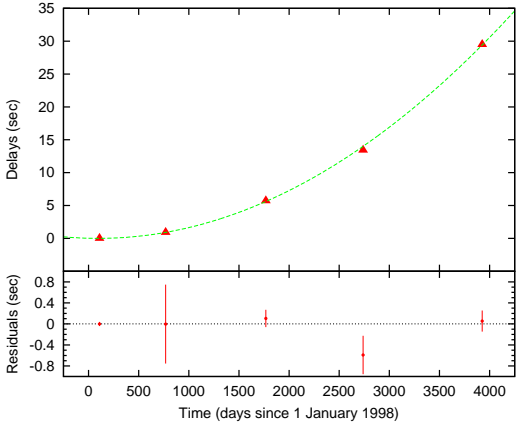}
\vspace{-1cm}
\caption{Differential correction to the time of passage of the neutron star
at the ascending node for each of the five outbursts analyzed (top panel) and
residuals with respect to the best fit parabola (bottom panel).}
\label{fig:fig4}
\end{figure}

In agreement with previous results, we find a highly significant derivative 
of the orbital period, which indicates that the orbital period in this 
system is increasing at a rate of $\dot  P_{\rm orb} = (3.89 \pm 0.15) \times 
10^{-12}$ s/s. The best-fit values for the orbital parameters are shown in 
Table~\ref{tabps}. 
With these new ephemerides, we correct the events of the {\xmm} ToO 
observation of 2008 October 1 and fit the 700 s folded phase delays of 
the fundamental with a straight line to obtain the correction to the
spin frequency adopted for the folding. The fit is acceptable 
with a reduced $\chi^2 = 1.16$ (for 71 d.o.f.).
The best-fit spin frequency of the 2008 outburst is also reported in 
Table~\ref{tabps}.

\section{Discussion}

In this paper we present a timing analysis of the {\xmm} observation performed
during the last outburst of the AMSP {\saxj}, together with a re-analysis of
the past four outbursts from this source observed by {\it RXTE}/PCA.
We find that the measures of the time of ascending node passage
for the last 5 outbursts from this source are again 
perfectly fitted by a parabola; from this fitting, we find a derivative of the 
orbital period, $\dot P_{\rm orb} = (3.89 \pm 0.15) \times 10^{-12}$ s/s, 
which agrees with previous measures (H08, DS08; see also
\citet{Patruno_08} and \citet{Hartman_09}, who report a 
similar value using {\it RXTE} data taken during the 2008 outburst). 
The conclusion from these measurements is that the orbital-period 
derivative in this system has been stable over the past ten years. 
This supports the hypothesis that the system is following a 
non-conservative secular binary evolution (see DS08).

\begin{table}
\begin{minipage}[t]{\columnwidth}
\caption{Best-fit orbital solution for {\saxj} derived from the analysis of
the five outbursts observed from 1998 to 2008.}
\label{tabps}
\centering
\renewcommand{\footnoterule}{}  
\begin{tabular}{llr}
\hline \hline
Parameter  & Units & Value  \\
\hline
$T^*_{1998}$ & MJD  		& $50914.79452952(85)$ \\
$P_{\rm orb}$  & s	& $7249.156444(23)$ \\
$\dot P_{orb}$ & $ 10^{-12}$ s/s  	& $3.89(15)$ \\
$a_1 \sin i / c$ & s  & $0.0628106(20)$ \\
$e$		&	& $< 1.2 \times 10^{-4}$ \\
$\nu_{2008}$  & Hz & $400.9752084532(17)$ \\
$T^*_{2008}$ & MJD & 54739.98342965 \\
\hline
\end{tabular} 
\end{minipage}
Note: 
Errors are at $90\%$ c.l. on the last 2 digits. The upper
limit on the eccentricity is at $95\%$ c.l.
The values of $P_{\rm orb}$ and $\dot P_{orb}$ are referred to $T^*_{1998}$,
while the value of $\nu_{2008}$ is referred to $T^*_{2008}$.
\end{table}

H08 note that the orbital parameters of {\saxj} are very similar to those of 
the so-called black-widow millisecond pulsars, which show large and variable 
$\dot P_{\rm orb}$ \citep[see e.g.][]{Arzoumanian_94,Doroshenko_01}. 
In those cases $\dot P_{\rm orb}$ changes have been ascribed to gravitational 
quadrupole coupling, i.e. a variable quadrupole moment of the companion 
that is caused by a cyclic spin-up and spin-down in the upper layers of 
the companion \citep{Applegate_92}. 
In this scenario, the companion star must be partially non-degenerate, 
convective, and magnetically active, so that the wind of the companion 
star will result in a strong torque, which tends to slow  the star down.
This torque is then transferred to the orbit via gravitational interaction.
Typical timescales for the orbital period (roughly sinusoidal) modulation
are about $5-6$~yr \citep{Applegate_94}.
However, {\saxj} has been monitored since its discovery as a millisecond pulsar 
in 1998 and does not show any evidence of this complex behavior in its
orbital period. Instead, its $\dot{P}_{\mathrm{orb}}$ has been
remarkably stable over this time. We cannot exclude, however, that its
$\dot{P}_{\mathrm{orb}}$ might exhibit more complex behavior on decade-long 
time scales.

The value of the orbital period derivative of {\saxj} clearly indicates that the binary
system is expanding at a rate that is more than one order of magnitude higher
than what is predicted by conservative mass transfer driven by GR (H08, DS08). 
In fact, the angular momentum losses caused by GR have the effect of reducing
the binary orbital period. 
The transfer of mass from the secondary to the neutron 
star indeed has  the effect of expanding the system (since the secondary star is
lighter than the neutron star, e.g. \citealt{Verbunt_93}). 
Following DS08, angular momentum conservation together with the third Kepler's law
and considering the GR angular momentum losses, give the orbital period derivative: 
\begin{eqnarray}
\label{dotP_saxj}
\dot P_{-12} = &0.138 & m_1^{5/3} q_{0.1} (1 + 0.1 q_{0.1})^{-1/3} P_{\rm 2h}^{-5/3} + 
\nonumber \\ 
&6.840 & m_1^{-1} q_{0.1}^{-1} P_{\rm 2h} \; g(\beta,q,\alpha)\; \dot m_{-9} 
\end{eqnarray}
where $\dot P_{-12}$ is $\dot P_{\rm orb}$ in units of $10^{-12}$ s/s,
$ m_1 = M_1/M_\odot$,
$ q_{0.1}$ is the mass ratio $q = m_2 / m_1$ in units of 0.1,
$P_{\rm 2h} =  P_{\rm orb}/2{\rm h}$,
$\dot m_{-9} = - \dot M_2/(10^{-9} M_\odot/{\rm yr})$,
$\beta$ is the fraction of the mass lost by the companion that is 
accreting onto the neutron star,
$\alpha$  the specific angular momentum of the mass lost by the system 
written in units of the specific angular momentum of the companion 
($\alpha = [1-0.462 (1+q)^{2/3} q^{1/3}]^2 \simeq 0.7$ for matter leaving 
the system with the specific angular momentum of the inner Lagrangian point),
and $g(\beta,q,\alpha) = 1 - \beta q - (1-\beta) (\alpha + q/3)/(1+q)$.
Inserting our derived value $\dot P_{-12} = 3.89$ into this equation, we can 
derive $\dot m_{-9}$ vs.\ $m_2 = q m_1$. 
The corresponding curves (very insensitive to the adopted values 
$m_1 = 1.1, 1.56, 2.2$)
are shown in Fig.~\ref{fig:fig5} and labeled with $\beta = 1$ for the 
conservative mass transfer case.  
\begin{figure}
\hspace{-1.5cm}
\includegraphics[angle=-90,width=10.0cm]{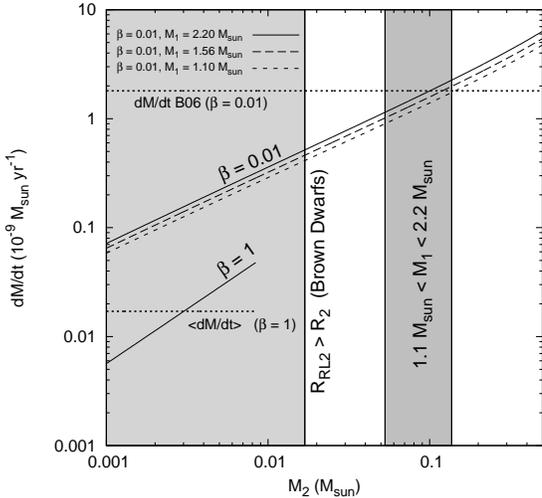}
\vspace{-0.5cm}
\caption{Mass-loss rate of the companion of {\saxj} in conservative ($\beta =1$)
and highly non-conservative ($\beta = 0.01$) scenarios. The gray area on the left  
is forbidden because the radius of the (Brown Dwarf) companion is greater than its
Roche Lobe. The gray area on the right shows the constraints on the companion mass 
imposed by the mass function of the system.}
\label{fig:fig5}
\end{figure}
In the same figure the horizontal line at $\dot m_{-9} = 0.017$ is the time-averaged 
mass accretion rate derived from the X-ray luminosity observed 
during outbursts and quiescence periods in the last ten years
(which could even be an overestimate by a factor 2 of 
the actual mass accretion rate, cf.\ \citealt{Galloway_06}). 
The crossing point of these curves, which is the {\it necessary solution 
imposed by the conservation of angular momentum}, falls in the forbidden 
region at $m_2 \sim 0.003$.

We can further impose the obvious constraint 
$R_2 \leq R_{\rm RL_2}$, where $R_2$ is the radius of the companion and 
$ R_{\rm RL_2} = 1.2 \times 10^{10} m_2^{1/3} P_{\rm 2h}^{2/3}$ cm is
the radius of the Roche Lobe of the companion.
Adopting results from detailed numerical calculations of the mass radius 
relation for low main sequence stars and brown dwarfs of minimum radius
(namely cold, or older than 5 Gyr, \citealt{ChabrierBaraffe_00}), 
the gray area on the left in Fig.~\ref{fig:fig5} 
indicates the region excluded by the violation of this constraint and shows 
that a conservative solution is impossible. Companion masses 
below $0.02 \; M_\odot$ are also excluded by the constraints imposed by
the mass function of the system for reasonable values of the neutron star mass.

In line with DS08, we thus propose a highly non-conservative
scenario in which the mass-loss rate from the companion is stable
at a level of $\dot m_{-9} \sim 1$,
with accretion episodes lasting a few tens of days, separated by quiescent phases 
in which {\it the same rate of mass is ejected by the system}. In this hypothesis,
we calculate that $\beta = 9.5 \times 10^{-3}$. 
The corresponding lines (which are again computed for $m_1 = 1.1, 1.56, 2.2$)
are shown in Fig.~\ref{fig:fig5} and labeled with $\beta = 0.01$.
The crossing point between the non-conservative lines and the horizontal line 
for $\dot m_{-9} = 1.8$ (from B06) at $m_2 \sim 0.08$ is outside the forbidden 
area and demonstrates that our proposed scenario is, in principle, viable.
The ejection of matter out of the system via a pulsar wind will not
imply an extra spin down for the pulsar, since in our model matter is
ejected by the radiation pressure of the pulsar, well outside the light
cylinder radius (where the electromagnetic radiation from the pulsar is no
longer connected to the neutron star magnetic field). 
Also, any possible spin down caused by the propeller effect at the end of
an outburst may be averaged to zero by a spin up, because of the accretion
inside the corotation radius, that is expected at the beginning of an outburst.

In conclusion, to explain the large {\it and stable} orbital period derivative
in the case of {\saxj}, we propose that mass is expelled  at a high rate 
from the system with the specific angular momentum of the inner Lagrangian 
point (or, eventually, of an outer disk, if present, e.g.\ \citealt{Deloye_08}).
These results strengthen the hypothesis  that {\saxj} belongs to the population 
of the so-called `hidden' millisecond pulsars, whose radio emission is completely
blocked by material engulfing the system that is continuously replenished
by the mass outflow driven by companion irradiation. 
In the case of {\saxj}, irradiation is probably caused by 
the power emitted by the magneto-dipole rotator, which may also explain why,
although the companion star is transferring mass at a high rate, this mass
does not accrete onto the neutron star (see DS08 for a more detailed 
discussion). If this is the case, $\sim 99\%$ of the transferred mass
in this system is not directly observable.
Although 10 years is not long enough to confirm that the system
is in a stationary (secular) orbital evolution, the next few years will be
important for confirming the stability of the orbital period 
derivative or to observe $\dot P$ reversal as in classical black-widow pulsars. 

\begin{acknowledgements}
We thank the {\xmm} team who supported this ToO observation,
and the referee for useful discussions.
\end{acknowledgements}

\bibliographystyle{aa}
\bibliography{Bibliografia}

\end{document}